\documentclass{iau}
\usepackage{graphicx,natbib}
 
\newcommand{\apj}{ApJ}           
\newcommand{\apjl}{ApJ}           
\newcommand{\mnras}{MNRAS}       

\newcommand{\aap}{A\&A}

\newcommand{\aj}{AJ}
\newcommand{\pasp}{PASP}
\newcommand{\apjs}{ApJS}           

\title{IFUs surveys, a panoramic view of galaxy evolution}

\author[S\'anchez]{Sebasti\'an F. S\'anchez$^1$ \and The CALIFA collaboration}

\affiliation{$^1$Instituto de Astronom\'\i a,Universidad Nacional Auton\'oma de Mexico, A.P. 70-264, 04510, M\'exico,D.F. email: {\tt sfsanchez@astro.unam.mx}}

\pubyear{2014}
\volume{309}
\jname{Galaxies in 3D across the Universe}
\editors{B. L. Ziegler, F. Combes, H. Dannerbauer, M. Verdugo, eds.}

\begin{document}

\maketitle

\begin{abstract}
We present here a brief summary of the currenly on-going IFU surveys of galaxies in teh Local Universe,
describing their main characteristics, including their sample
selections, instrumental setups, wavelength ranges, and area of the
galaxies covered. Finally, we make an emphasis on the main characteristics
of the CALIFA survey and the more recent results that has been recently
published.
\end{abstract}

\firstsection
\section{Introduction}

Much of our recently acquired understanding of the architecture of the
Universe and its constituents derives from large surveys (e.g.,
2dFGRS, Folkes et al. 1999;SDSS, York et al. 2000; GEMS, Rix et
al. 2004; VVDS, Le F{\`e}vre et al. 2004; COSMOS, Scoville et al. 2007
; GAMMA, Driver et al. 2009, to name but a few). Such surveys have not
only constrained the evolution of global quantities such as the cosmic
star formation rate, but also enabled us to link this with the
properties of individual galaxies -- morphological types, stellar
masses, metallicities, etc.. Compared to previous approaches, the
major advantages of this recent generation of surveys are: (1) the
large number of objects sampled, allowing for meaningful statistical
analysis to be performed on an unprecedented scale; (2) the
possibility to construct large comparison/control samples for each
subset of galaxies; (3) a broad coverage of galaxy subtypes and
environmental conditions, allowing for the derivation of universal
conclusions; and (4) the homogeneity of the data acquisition,
reduction and (in some cases) analysis.

On the other hand, the cost of these surveys, in terms of telescope time,
manpower, and involved time scales, is also unprecedented in astronomy. The
user of such data products has not necessarily been involved in any step of
designing or conducting the survey, but nevertheless takes advantage of the
data by exploiting them according to her/his scientific interests.  This new
approch to observational astronomy is also changing our perception of the
scientific rationale behind a new survey: While it is clear that certain
planned scientific applications are key determinants to the design of the
observations and `drive' the survey, the survey data should at the same time
allow for a broad range of scientific exploitation. This aspect is now often
called {\it Legacy} value.

Current technology generally leads to surveys either in the imaging or in the
spectroscopic domain.  While imaging surveys provide two-dimensional coverage,
they carry very little spectral information. This is also true for 
 multiband photometric surveys such as COMBO-17 (Wolf et
al. 2003), ALHAMBRA (Moles et al$.$ 2008) or the planned PAU project
(Ben\'{\i}tez et al$.$ 2009), which will still be unable to accurately capture
individual spectral lines and measure, e.g., emission line ratios or internal
radial velocity differences. Spectroscopic surveys such as SDSS or zCOSMOS, on
the other hand, do provide more detailed astrophysical information, but they
are generally limited to one spectrum per galaxy, often with aperture losses
that are difficult to control. For example, the $3''$ diameter of the fiber
used in the SDSS corresponds to vastly different linear scales at different
redshifts, without the possibility to correct for these aperture effects.

An observational technique combining the advantages of imaging and
spectroscopy (albeit with usually quite small field of view) is
Integral Field Spectroscopy (IFS). However, so far this technique has
rarely been used in a `survey mode' to investigate large
samples. Among the few exceptions there is, most notably, the SAURON
survey (de Zeeuw et al. 2002), focused on the study of the central
regions of 72 nearby early-type galaxies and bulges of spirals. Others
are (i) the PINGS project at the CAHA 3.5m of a dozen very nearby
galaxies (Rosales-Ortega et al. 2010), (ii) the Disk Mass Survey
(Bershady et al. 2010) that provided super high spectral resolution
using PPAK and SparsePak of 46 face-on spirals; (iii) the VENGA
project (Blanc et al. 2013), that is observing a sample of 32 face-on
spirals with VIRUS-P; and (iv) and the SIRIUS project, currently
studying 70 (U)LIRGS at $z<0.26$ using different IFUs (Arribas et
al. 2008), among others.  However, despite the dramatic improvement
over previous data provided by these `surveys', they are all affected
by non-trivial sample selection criteria and, most importantly,
incomplete coverage of the full extent of the galaxies.

In the last few years this situation has changed dramatically, and it
will change even more in the upcomming years, with the development of
IFS surveys of galaxies, in particular Atlas3D (Cappellari et al. 2011
), CALIFA (S\'anchez et al. 2012), MaNGA (Law et al. 2014) and SAMI
(Croom et al. 2012). We review here the current on-going or recently
started surveys, making an emphasis on the result obtained by the
CALIFA survey (S\'anchez et al. 2012), the first of those surveys
providing wih a panoramic view of galaxy properties.


\begin{table}
\caption{Comparison of IFU Surveys}
\label{tab:comp}      
\begin{center}
\begin{tabular}{lcccc}        
\hline\hline                 
 Specification & MaNGA & SAMI & CALIFA & Atlas3D \\
\hline
Sample Size                        & 10,000 & 3,400 & 600 & 260 \\
Selection                          & $M>10^9 M_{\cdot}$ & $M>10^{8.2} M_{\cdot}$ & 45$" < D_{25} < 80 "$ & E/S0$M>10^{9.8} M_{\cdot}$ \\
Radial coverage                    & 1.5$r_e$ (2/3), 2.5$r_e$ (1/3)& 1-2$r_e$ & $>$2.5$r_e$& $<$1$r_e$ \\
S/N at 1$r_e$                      & 15-30& 10-30 & $\sim$30 & 15\\
Wavelength range(\AA)              & 3600-10300 & 3700-7350 & 3700-7500& 4800-5380\\
Instrumental resolution  & 50-80 km/s& 75/28 km/s& 85/150 km/s&  105 km/s\\
Input Spaxel Size                  & 2.0$''$& 1.6'$''$& 2.7$''$& 1$''$\\
Input Spaxels per object           & $<$3$\times$127$^1$ & 3$\times$61 & 3$\times$331 & 1,431 \\
Spatial FWHM                       & 2$''$ & 2$''$& 2.5$''$ & 1.5$''$ \\
Telescope size                     & 2.5m  & 3.5m & 3.5m & 4.2m \\
\hline                                   
\end{tabular}

$^1$ This corresponds to the largest MaNGA bundle. Each MaNGA plate provides with 17 bundles of different amount of fibers: 2x19; 4x37; 4x61; 2x91; 5x127.

\end{center}
\end{table}

\section{Major recent and on-going surveys}

Most of the listed on-going IFU surveys in the Local Universe have as
a major goal to understand the nature of galaxies as a consequence of
the evolution through cosmologocial times. Therefore, most of their
differences are on the technical details of the selected IFU, the
adopted setup and the sample selection. In general, all the surveys
tries to select a sample representative of the population of galaxies
at low redshifts. Therefore, most of them cover mostly all galaxy
types, covering as much as possible the color-magnitude
diagram. However the adopt different selection criteria.  In the case
of MaNGA and SAMI the main criteria was to cover a wide range of
galaxy masses. In MaNGA it was forced that the distribution along
masses was as flat as possible. SAMI adopted a different procedure,
selecting a set of volume-limited sub-samples, together with a set of
sub-samples centred in galaxy clusters. In the previous case the
sample was derived from the SDSS catalogs, while in the former one it
was extracted from the GAMMA survey. The CALIFA sample selection is
based on their optical size, i.e., it is a diameter selection that
guaranteed that all galaxies are observed in most of their optical
extension, as we explain below. Atlas3D adopted a different scheme,
their main science goal (to understand the building up of bulges and
early-type galaxies). The selected early-type/red galaxies from a
volume-limited sample up to $\sim$42 Mpc (being the sample at lower
redshiffs of the different surveys).

\begin{figure*}
\centering
\includegraphics[width=1.0\columnwidth]{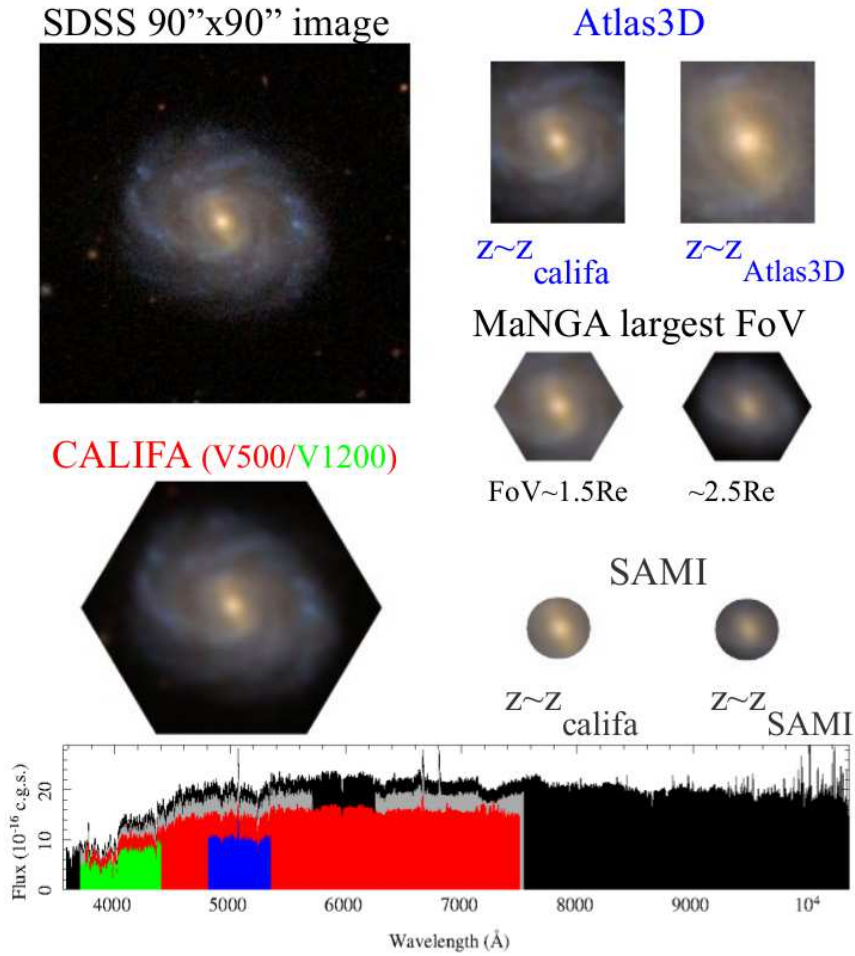} 
\caption{Schematic figure comparing the different on-going major IFU surveys. The top-left figure shows a 90$"$$\times$90$"$ post-stamp image of NGC5947 extracted from the SDSS imaging survey. The same image as it would be observed by the four different IFU surveys discussed in this article are presented in the different post-stamp figures. For MaNGA we have selected the bundle with the largest FoV, although we should remind the reader that only 1/5th of the survey will be observed using this bundle. Finally, the wavelength range covered by each survey is presented in the bottom panel. We should remind that the spectral resolution is different for each survey. For SAMI the largest resolution is achieved in the redder wavelength range covered by this survey.}\label{fig:comp}
\end{figure*}

Figure \ref{fig:comp} and Table \ref{tab:comp} presents an schematic
comparison between the different surveys. The differences between
different surveys are clearly highlighted in there.  While Atlas3D
provides the highest spatial resolution and better spatial sampling
for the individual galaxies, if offers a limited wavelength range and
FoV compared to the size of the galaxies. In other cases, like MaNGA,
the total number of objects is the main statistical advantage, since
it would provide with the average characteristics properties of the
main population of galaxies for many different galaxy types with
better statistical significance than any of the other surveys. The
penalty is that not all the galaxies will be sampled by the same
number of fibers (ranging between 3x19 to 3x127), and therefore the
physical sampling is different for different galaxies. In the case of
SAMI the main adventage is the wider range of galaxy masses sampled
and the higher spectral resolution in the red wavelength
range. However the large redshift range implies a different physical
sampling of the different galaxies. Finally, in the case of CALIFA, it
provides the widest spatial coverage compared to the spatial size of
the galaxies, and one of the best spatial physical
resolutions. However, the sample is more limited than the one studiest
by MaNGA or SAMI. In summary we consider that each of the IFUs surveys
is very complementary.

\section{CALIFA: A brief introduction}

The Calar Alto Legacy Integral Field Area (CALIFA) survey (Sánchez et
al. 2012a) is an ongoing large project of the Centro Astronómico
Hispano-Alemán at the Calar Alto observatory to obtain spatially
resolved spectra for 600 local (0.005$<z<$0.03) galaxies by means of
integral field spectroscopy (IFS). CALIFA observations started in June
2010 with the Potsdam Multi Aperture Spectrograph (PMAS, Roth et
al. 2005), mounted to the 3.5 m telescope, utilizing the large
(74$"$$\times$64$"$) hexagonal field-of-view (FoV) offered by the PPak
fiber bundle (Verheijen et al. 2004; Kelz et al. 2006). PPak was
created for the Disk Mass Survey (Bershady et al. 2010). Each galaxy
is observed using two different setups, an intermediate spectral
resolution one (V1200, $R\sim 1650$), that cover the blue range of the
optical wavelength range (3700-4700\AA), and a low-resolution one
(V500,$R\sim 850$, that covers the first octave of the optical
wavelength range (3750-7500\AA). A diameter-selected sample of 939
galaxies were drawn from the 7th data release of the Sloan Digital Sky
Survey (SDSS, Abazajian et al. 2009) which is described in Walcher et
al. (2014). From this mother sample the 600 target galaxies are
randomly selected.

Combining the techniques of imaging and spectroscopy through optical
IFS provides a more comprehensive view of individual galaxy properties
than any traditional survey. CALIFA-like observations were collected
during the feasibility studies (Mármol-Queraltó et al. 2011; Viironen
et al. 2012) and the PPak IFS Nearby Galaxy Survey (PINGS,
Rosales-Ortega et al. 2010), a predecessor of this survey. First
results based on those datasets already explored their information
content (e.g. Sánchez et al. 2011; Rosales-Ortega et al. 2011;
Alonso-Herrero et al. 2012; Sánchez et al. 2012b; Rosales-Ortega et
al. 2012). CALIFA can therefore be expected to make a substantial
contribution to our understanding of galaxy evolution in various
aspects including, (i) the relative importance and consequences of
merging and secular processes; (ii) the evolution of galaxies across
the color–magnitude diagram; (iii) the effects of the environment on
galaxies; (iv) the AGN-host galaxy connection; (v) the internal
dynamical processes in galaxies; and (vi) the global and spatially
resolved star formation history of various galaxy types.

Compared with other IFS surveys, CALIFA offers an unique combination
of (i) a sample covering a wide range of morphological types in a wide
range of masses, sampling the Color-Magnitude diagram for M$_g>-$ 18
mag; (ii) a large FoV, that guarantees to cover the entire optical
extension of the galaxies up to 2.5$r_e$ for an 80\% of the sample;
and (iii) an accurate spatial sampling, with a typical spatial
resolution of $\sim$1 kpc for the entire sample, which allows to
optical spatial resolved spectroscopic properties of most relevant
structures in galaxies (spiral arms, bars, buges, Hii regions...). The
penalty for a better spatial sampling of the galaxies is the somehow
limited number of galaxies in the survey (e.g., MaNGA and SAMI). In terms of the spectral
resolution, only the blue wavelength range is sampled with a similar
spectral resolution than these two other surveys.

As a legacy survey, one of the main goals of the CALIFA collaboration
is to grant public access of the fully reduced datacubes. In November
2012 we deliver our 1st Data Release (Husemann et al. 2013),
comprising 200 datacubes corresponding to 100
objects \footnote{http://califa.caha.es/DR1/}. After almost two years,
and a major improvement in the data reduction, we present our 2nd Data Release
(Garcia Benito et al., in prep.), comprising 400 datacubes corresponding to 200
objects \footnote{http://califa.caha.es/DR2/}, the 1st of October 2014.

\begin{figure*}
\centering
\includegraphics[width=1.0\columnwidth]{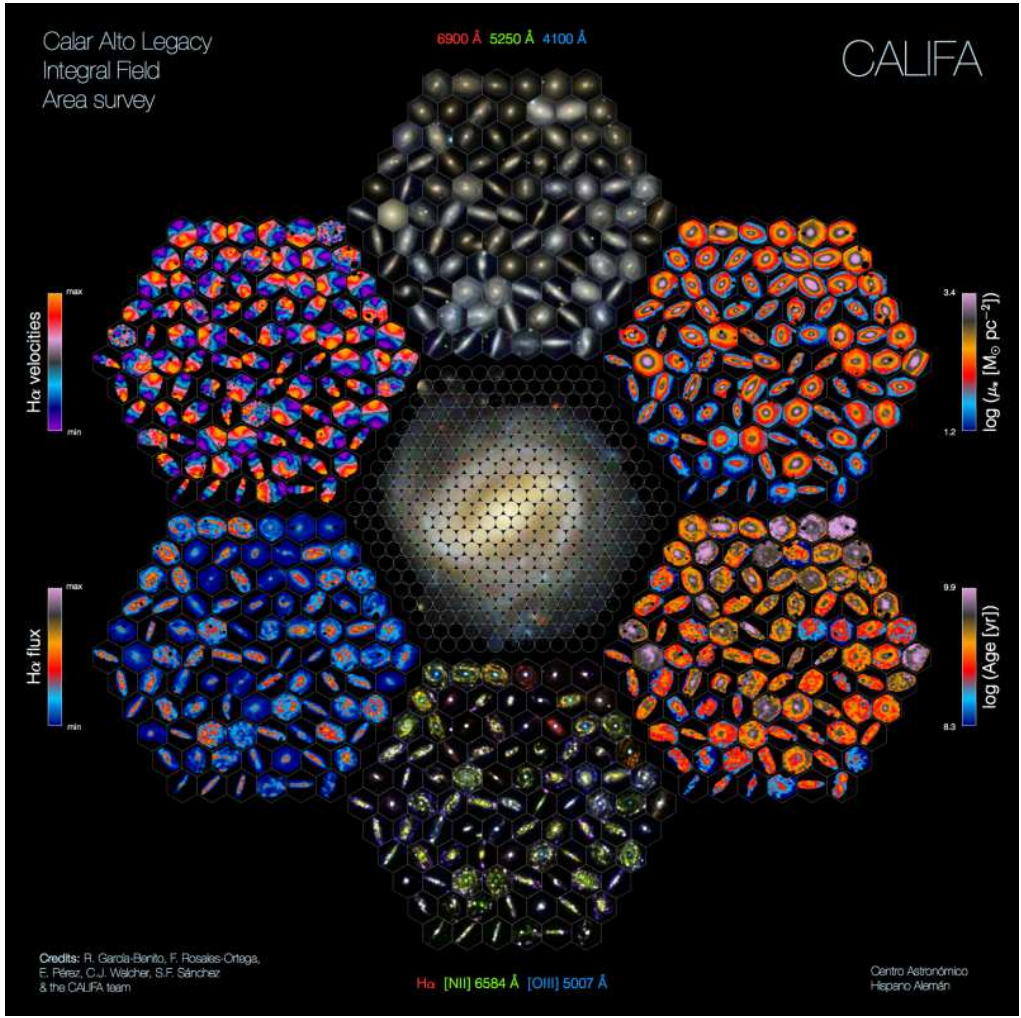} 
\caption{Schematic representation of the different dataproducts that can be obtained for sample of galaxies with the state-of-the art IFU surveys. In this case it shows for a set of 100 galaxies extracted from the CALIFA survey six hexagons showing (following the clock arrows): (i) The true-color maps showing the light distribution; (ii) The stellar mass density distribution; (iii) The distribution of the luminosity weighted ages across the galaxies; (iv) a three-color image showing the ionized gas distribution for H$\alpha$, [NII]$\lambda$6584 and [OIII]$\lambda$5007; (v) the H$\alpha$ intensity distribution and (vi) the ionized gas kinematics. Similar dataproducts are derived for Altas3D, MaNGA and SAMI datasets. }\label{fig:hexa}
\end{figure*}

\section{CALIFA: Main Science Results}

Figure \ref{fig:hexa} illustrate the dataproducts that can be derived
from the IFU datasets obtained by the CALIFA survey, comprising a
panoramic view of the spatial resolved spectroscopic prorperties of
these galaxies, including information of the stellar populations,
ionized gas, mass distribution and stellar and gas kinematics. Similar
dataproducts are derived for any of the indicated projects: Atlas3D,
MaNGA or SAMI.

Different science goals have been already addressed using this
information: (i) New techniques has been developed to understand the
spatially resolved star formation histories (SFH) of galaxies (Cid
Fernandes et al., 2013, 2014). We found the solid evidence that
mass-assembly in the typical galaxies happens from inside-out (P\'erez
et al., 2013). The SFH of bulges and early-type galaxies are
fundamentally related to the total stellar mass, while for disk
galaxies it is more related to the local stellar mass density
(Gonz\'alez Delgado et al., 2013 \& 2014); (ii) We developed new tools
to detect and extract the spectroscopic information of HII regions
(S\'anchez et al., 2012b), building the largest catalog currently
available ($\sim$6,000 HII regions and aggregations). This catalog has
been used to define a new oxygen abundance calibrator anchored with
electron temperature measurements (Marino et al., 2013). From these,
we explored the proposed dependence of the Mass-Metallicity relation
with SFR (S\'anchez et al., 2013), and the local Mass-Metallicity
relation (Rosales-Ortega et al. 2012). We found that all galaxies in
our sample present a common abundance radial gradient with a similar
slope when normalized to the effective radius (S\'anchez et al.,
2014), which agrees with the proposed inside-out scenario for galaxy
growth. This characteristic slope is independent of the properties of
the galaxies, and in particular of the presence or absence of a bar,
contrary to previous results. More recently, this result has been
confirmed by the analysis of the stellar abundance gradient in the
same sample (S\'anchez-Bl\'azquez et al., 2014). We also explore the
progenitors of SuperNovae based on the gas phase metallicity and loca
SFR at the location of the explosion (Gabany et al. 2014); (iii) We
explore the origin of the low intensity, LINER-like, ionized gas in
galaxies. These regions regions are clearly not related to
star-formation activity, or to AGN activity. They are most probably
relatd to post-AGB ionization in many cases (Kehrig et al., 2012;
Singh et al., 2013; Papaderos et al. 2013); (iv) We explore the
aperture and resolution effects on the data. CALIFA provides a unique
tool to understand the aperture and resolution effects in larger
single-fiber (like SDSS) and IFS surveys (like MaNGA, SAMI). We
explored the effects of the dilution of the signal in different gas
and stellar population properties (Mast et al., 2014), and proposed an
new empirical aperture correction for the SDSS data (Iglesias-P\'aramo
et al., 2013); (v) CALIFA is the first IFU survey that allows gas and
stellar kinematic studies for all morphologies with enough
spectroscopic resolution to study (a) the kinematics of the ionized
gas (Garc\'ia-Lorenzo et al., 2014), (b) the effects of bars in the
kinematics of galaxies (Barrera-Ballesteros et al. 2014a); (c) the
effects of the intraction stage on the kinematic signatures
(Barrera-Ballesteros et al., submitted), (d) measure the Bar Pattern
Speeds in late-type galaxies (Aguerri et al., submitted), (iv) extend
the measurements of the angular momentum of galaxies to previously
unexplored ranges of morphology and ellipticity (Falc\'on-Barroso et
al., in prep.); and (v) finally we explore in detail the effects of
galaxy interaction in the enhancement of star-formatio rate and the
ignition of galactic outflows (Wild et al. 2014).

\section{Conclusions}

Along this review we have presented the main characteristics of IFU
surveys in the Local Universe, describing their main advantage with
respect to precedent techniques (like single fiber spectroscopic
surveys or imaging surveys). We made a summary of some of the first
attempts of performing a major IFU survey, and we describe the current
on-going or recently finished ones. Finally, we present the main
data-products that can be derived using these surveys and made a brief
summary of the science goals achieved recently by the CALIFA survey,
one of the most advanced on-going surveys that sample the full range
of morphological types of galaxies.

Recent results already presented based on the pilot and even the main
samples of the recently started MaNGA and SAMI surveys in this
conference, and the quality shown by MUSE data (Bacon et al., ),
illustrate how IFU surveys would fundamental tools in our understanding
of the evolution of galaxies.

\section*{Acknowledgements}

\noindent

SFS thank the support of the Mexican CONACyt Grant 180125.

We thank K. Bundy, S. Croom, M. Bershady, M. Cappellari for providing us
with fundamental information and useful comments that has helped in the redaction
of this document.

We thank the full CALIFA collaboration for providing us with most of the information
on the results from the survey, and in particular to R. Garcia Benito for providing
us with Fig. 3.

\end{document}